\documentclass[aps,prl,superscriptaddress,twocolumn,showpacs]{revtex4}
\usepackage{graphics}
\usepackage{amsmath}
\usepackage{graphicx}
\usepackage{amsfonts}
\usepackage{amssymb}

\begin{document}

\title{Characterization of collective Gaussian attacks and security of
coherent-state quantum cryptography}
\author{Stefano Pirandola}
\affiliation{MIT - Research Laboratory of Electronics, Cambridge MA 02139, USA}
\author{Samuel L. Braunstein}
\date{\today }
\affiliation{Computer Science, University of York, York YO10 5DD, United Kingdom}
\author{Seth Lloyd}
\affiliation{MIT - Research Laboratory of Electronics, Cambridge MA 02139, USA}
\affiliation{MIT - Department of Mechanical Engineering, Cambridge MA 02139, USA}
\date{\today }

\begin{abstract}
We provide a simple description of the most general collective Gaussian
attack in continuous variable quantum cryptography. In the scenario of such
general attacks, we analyze the asymptotic secret-key rates which are
achievable with coherent states, joint measurements of the quadratures and
one-way classical communication.
\end{abstract}

\maketitle

During recent years, quantum systems with infinite dimensional Hilbert
spaces have become the object of increasing interest within the quantum
information community \cite{CVbook}. These systems are generally called
\emph{continuous variable} systems and their standard prototype is given by
the bosonic modes of the radiation field. In ordinary experiments in quantum
optics, bosonic modes are generated in states with Gaussian statistics \cite%
{EisertGauss}, and these statistics are commonly preserved during subsequent
optical manipulation. Further, the Gaussian statistics can be preserved at
the end of quantum communication lines (e.g., optical fibers), where noisy
transformations of the state are induced by the interaction with an external
environment. From a theoretical point of view, the standard model for this
kind of transformation is represented by the one-mode Gaussian channel. This
is a completely positive trace-preserving (CPT) map that transforms Gaussian
states into Gaussian states, without creating any kind of correlation among
the various bosonic modes. The mathematical structure of this map is
relatively simple and has been further simplified in Ref.~\cite%
{HolevoCanonical} via the introduction of canonical forms.

In the context of continuous variable quantum key distribution (cvQKD),
one-mode Gaussian channels can be interpreted as the effect of collective
Gaussian attacks. Starting from this consideration, here we extend the
results of Refs.~\cite{HolevoCanonical,HolevoVittorio} to provide a full
characterization of the most general collective Gaussian attack in cvQKD.
Recall that collective Gaussian attacks have been recognized as the most
powerful collective attacks in cvQKD with Gaussian resources \cite{AcinNew}.
Furthermore, under suitable conditions \cite{Renner2}, collective attacks
have been recently proven to bound the most general attacks (coherent
attacks) against cvQKD protocols. Using our general characterization of
collective Gaussian attacks, we then analyze the security of a cvQKD
protocol, where coherent states are used to generate secret correlations.
Such a protocol is a simple generalization of the non-switching protocol of
Ref.~\cite{Hetero}, where further post-processing of the classical data is
also used to compensate possible squeezing and rotation of the output
quadratures.

\begin{figure}[tbph]
\vspace{-0.0cm}
\par
\begin{center}
\includegraphics[width=0.5\textwidth] {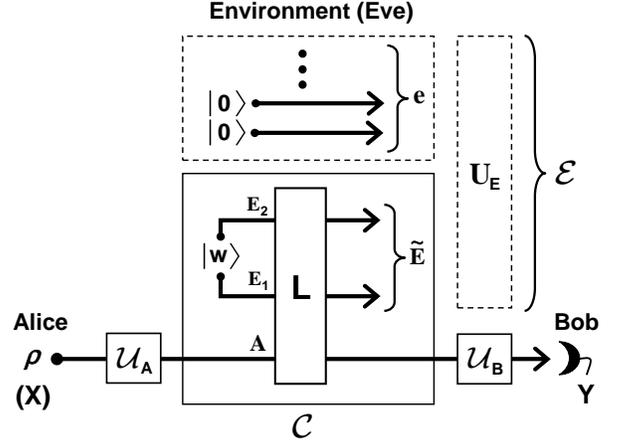}
\end{center}
\par
\vspace{-0.6cm}
\caption{The general scenario in five steps. \textbf{(1)~Quantum
communication.} \ Alice randomly picks signal states $\protect\rho $ from an
ensemble $\mathcal{A}$ encoding a classical variable $X$. At the output of
the channel, Bob detects the states via a quantum measurement. The
corresponding outcomes define an output classical variable $Y$ correlated to
$X$. \textbf{(2)~One-mode Gaussian channel.} A one-mode Gaussian channel $%
\mathcal{G}$ corresponds to a canonical form $\mathcal{C}$ up to a pair of
Gaussian unitaries $\mathcal{U}_{A}$ (at the input) and $\mathcal{U}_{B}$
(at the output). The central canonical form $\mathcal{C}$ can be dilated to
a symplectic interaction $\mathbf{L}$ involving two ancillary modes $\mathbf{%
E}:=\{E_{1},E_{2}\}$ prepared in a TMSV state $\left\vert w\right\rangle _{%
\mathbf{E}}$. The dilation of the form is unique up to isometries acting on $%
\mathbf{\tilde{E}}:=\{\tilde{E}_{1},\tilde{E}_{2}\}$. \textbf{(3)~Maximal
dilation.} By assuming Eve is in a finite box, the dilation can be extended
(via an identity) to the remaining modes $\mathbf{e}=\{e_{i}\}_{i=1}^{\infty
}$\ of the environment (prepared in vacua). This maximal dilation of $%
\mathcal{C}$\ is now unique up to unitaries $\hat{U}_{E}$ acting on $\{%
\mathbf{\tilde{E},e}\}$. \textbf{(4)~Collective Gaussian attack}. All the
output ancillas $\{\mathbf{\tilde{E},e}\}$ provide an ensemble $\mathcal{E}$%
, which Eve can detect to estimate $X$ or $Y$. By using an entropic bound
for Eve's accessible information, the extra ancillas and the extra unitary
(dashed boxes in the figure) can be neglected. As a consequence, only the
set $G:=\{\mathbf{L}(\protect\tau ,r),\left\vert w\right\rangle ,\mathcal{U}%
_{A},\mathcal{U}_{B}\}$ (solid boxes in the figure)\ is needed to
characterize the attack. \textbf{(5)~Coherent-state protocol}. Alice's
signal states $\protect\rho $\ are coherent states $\left\vert \protect%
\alpha \right\rangle $ whose amplitudes encode a Gaussian variable ($X=%
\protect\alpha $). Bob's measurement is a heterodyne detection retrieving
the output amplitudes ($Y=\protect\beta $).}
\label{PRLpic1}
\end{figure}

Let us consider a single bosonic mode, whose quadratures $\mathbf{\hat{x}}%
^{T}:=(\hat{q},\hat{p})$ satisfy $[\mathbf{\hat{x}},\mathbf{\hat{x}}^{T}]=2i%
\mathbf{\Omega }$, where the matrix $\mathbf{\Omega }$ is defined by the
entries $\mathbf{\Omega }_{11}=\mathbf{\Omega }_{22}=0$ and $\mathbf{\Omega }%
_{12}=-\mathbf{\Omega }_{21}=1$. Every Gaussian state $\rho $ \cite%
{EisertGauss} of the system is characterized by a displacement vector $%
\mathbf{\bar{x}}:=\mathrm{Tr}(\mathbf{\hat{x}}\rho )$ and a covariance
matrix $\mathbf{V}:=\mathrm{Tr}\{[\mathbf{\hat{x}\hat{x}}^{T}+(\mathbf{\hat{x%
}\hat{x}}^{T}\mathbf{)}^{T}]\rho \}/2-\mathbf{\bar{x}\bar{x}}^{T}$. In a
quantum communication scenario, this kind of state can be used by a sender
(Alice) to transmit classical information to a receiver (Bob) through a
noisy quantum channel. Usually, Alice chooses $\rho (\mathbf{\bar{x}},%
\mathbf{V})$ from an ensemble of $\emph{signal}$ states $\mathcal{A}:=\{p(%
\mathbf{\bar{x}}),\rho (\mathbf{\bar{x}},\mathbf{V})\}$ encoding a classical
variable $X:=\{p(\mathbf{\bar{x}}),\mathbf{\bar{x}}\}$. This variable
describes the modulation of the displacement $\mathbf{\bar{x}}$\ via some
probability distribution $p(\mathbf{\bar{x}})$. The signal states are then
sent to Bob, in independent uses of the quantum channel. At the output, Bob
gets a noisy ensemble $\mathcal{B}$, whose (incoherent) detection gives a
classical variable $Y$ which is correlated to $X$ (see Fig.~\ref{PRLpic1},
step~1). In this scenario, the standard model for the noise process is
represented by the one-mode Gaussian channel. By definition, this channel is
a CPT map $\mathcal{G}$ acting on a single bosonic mode and preserving the
Gaussian statistics of the input state. The mathematical description of this
channel is fully contained in a triplet $\{\mathbf{T},\mathbf{N,d}\}$, where
$\mathbf{d}$ is an $\mathbb{R}^{2}$ vector and $\mathbf{T},\mathbf{N}$ are $%
2\times 2$ real matrices \cite{Conditions}. Explicitly, the action of $%
\mathcal{G}(\mathbf{T},\mathbf{N,d})$\ on a Gaussian state $\rho (\mathbf{%
\bar{x},V})$ corresponds to the simple transformations%
\begin{equation}
\mathbf{\bar{x}\rightarrow \mathbf{T}\bar{x}+d~,~V}\rightarrow \mathbf{TVT}%
^{T}+\mathbf{N~.}  \label{Displ_Transf}
\end{equation}%
In particular, for $\mathbf{N=0}$ and $\mathbf{T}:=\mathbf{S}$ symplectic
(i.e., $\mathbf{S\Omega S}^{T}=\mathbf{\Omega }$), the channel represents a
Gaussian unitary. This means that we can set $\mathcal{G}(\mathbf{S},\mathbf{%
0,d}):=\mathcal{U}(\mathbf{S},\mathbf{d})$ where $\mathcal{U}:\rho
\rightarrow \hat{U}\rho \hat{U}^{\dagger }$ with $\hat{U}$ a unitary
operator.

Remarkably, the mathematical structure of $\mathcal{G}(\mathbf{T},\mathbf{N,d%
})$ can be further simplified thanks to recent results of Ref.~\cite%
{HolevoCanonical}. In fact, every $\mathcal{G}(\mathbf{T},\mathbf{N,d})$ can
be decomposed as $\mathcal{G}=\mathcal{U}_{B}\circ \mathcal{C}\circ \mathcal{%
U}_{A}$, where $\{\mathcal{U}_{A},\mathcal{U}_{B}\}$ are Gaussian unitaries,
while the map $\mathcal{C}$, called the \emph{canonical form}, represents a
Gaussian channel with $\mathbf{d}=\mathbf{0}$ and $\mathbf{T}_{c},\mathbf{N}%
_{c}$ diagonal. The explicit expressions of $\mathbf{T}_{c}$ and $\mathbf{N}%
_{c}$ depend on three symplectic invariants of the channel: the generalized
\emph{transmission} $\tau :=\det \mathbf{T}$ (ranging from $-\infty $ to $%
+\infty $), the \emph{rank} $r:=[$rk$(\mathbf{T})$rk$(\mathbf{N})]/2$ (with
possible values $r=0,1,2$) and the \emph{temperature }$\bar{n}$ (which is a
positive number related to $\det \mathbf{N}$ \cite{Thermal}). These three
invariants $\{\tau ,r,\bar{n}\}$ completely characterize the two matrices $%
\mathbf{T}_{c},\mathbf{N}_{c}$ and, therefore, the corresponding canonical
form $\mathcal{C}=\mathcal{C}(\tau ,r,\bar{n})$. In particular, the first
two invariants $\{\tau ,r\}$ determine the class of the form \cite%
{HolevoCanonical}. The full classification is explicitly shown in the
following table%
\begin{equation*}
\begin{tabular}{c|c||c||c||c|c}
$\tau $ & $~r~$ & Class & ~~~Form~~ & $\mathbf{T}_{c}$ & $\mathbf{N}_{c}$ \\
\hline
$0$ & $0$ & $A_{1}$ & $\mathcal{C}(0,0,\bar{n})$ & $\mathbf{0}$ & $(2\bar{n}%
+1)\mathbf{I}$ \\
$0$ & $1$ & $A_{2}$ & $\mathcal{C}(0,1,\bar{n})$ & $\frac{\mathbf{I}+\mathbf{%
Z}}{2}$ & $(2\bar{n}+1)\mathbf{I}$ \\
$1$ & $1$ & $B_{1}$ & $\mathcal{C}(1,1,0)$ & $\mathbf{I}$ & $\frac{\mathbf{I}%
-\mathbf{Z}}{2}$ \\
$1$ & $2$ & $B_{2}$ & $\mathcal{C}(1,2,\bar{n})$ & $\mathbf{I}$ & $\bar{n}%
\mathbf{I}$ \\
$1$ & $0$ & $B_{2}(Id)$ & $\mathcal{C}(1,0,0)$ & $\mathbf{I}$ & $\mathbf{0}$
\\
$(0,1)$ & $2$ & $C(Att)$ & $\mathcal{C}(\tau ,2,\bar{n})$ & $\sqrt{\tau }%
\mathbf{I}$ & $(1-\tau )(2\bar{n}+1)\mathbf{I}$ \\
$>1$ & $2$ & $C(Amp)$ & $\mathcal{C}(\tau ,2,\bar{n})$ & $\sqrt{\tau }%
\mathbf{I}$ & $(\tau -1)(2\bar{n}+1)\mathbf{I}$ \\
$<0$ & $2$ & $D$ & $\mathcal{C}(\tau ,2,\bar{n})$ & $\sqrt{-\tau }\mathbf{Z}$
& $(1-\tau )(2\bar{n}+1)\mathbf{I}$%
\end{tabular}%
\end{equation*}%
In this table, the values of $\{\tau ,r\}$ in the first two columns specify
a particular class $A_{1},A_{2},B_{1},B_{2},C$ and $D$ \cite{Table}. Within
each class, the possible canonical forms are expressed in the third column,
where also the third invariant $\bar{n}$ must be considered. The
corresponding expressions of $\mathbf{T}_{c},\mathbf{N}_{c}$ are shown in
the last two columns, where $\mathbf{Z}:=\mathrm{diag}(1,-1)$, $\mathbf{I}:=%
\mathrm{diag}(1,1)$ and $\mathbf{0}$ is the zero matrix.

Thus, an arbitrary one-mode Gaussian channel $\mathcal{G}(\mathbf{T},\mathbf{%
N,d})$ can be expressed by a unique canonical form $\mathcal{C}(\tau ,r,\bar{%
n})$ up to a pair of input-output Gaussian unitaries $\{\mathcal{U}_{A},%
\mathcal{U}_{B}\}$. Now, it is known that every quantum channel can be
represented by a unitary interaction coupling the signal system to an
environment, prepared in some initial state $\rho _{\mathbf{E}}$. When $\rho
_{\mathbf{E}}$ is pure, such a dilation is called a \textquotedblleft
Stinespring dilation\textquotedblright\ and is unique up to partial
isometries \cite{Stines}. By extending the results of Ref.~\cite%
{HolevoVittorio}, we easily construct the Stinespring dilations of all the
canonical forms. In detail, a generic $\mathcal{C}(\tau ,r,\bar{n})$ can be
dilated to a three-mode Gaussian unitary corresponding to a symplectic
transformation $\mathbf{L}=\mathbf{L}(\tau ,r)$ \cite{Decomposition}. This
transformation mixes the input state $\rho _{A}$ with a two-mode squeezed
vacuum (TMSV) state $\left\vert w\right\rangle _{\mathbf{E}}$ of variance $%
w=2\bar{n}+1$ (see Fig.~\ref{PRLpic1}, step~2). Compactly, we denote by $\{%
\mathbf{L}(\tau ,r),\left\vert w\right\rangle \}$ the Stinespring dilation
of a generic canonical form $\mathcal{C}(\tau ,r,\bar{n})$. For particular
choices of the class $\{\tau ,r\}$, this dilation corresponds to well-known
Gaussian models of interaction. In particular, for $\{\tau ,r\}=\{1,2\}$, it
corresponds to a universal Gaussian cloner \cite{Cerf}, while for $0<\tau <1$
and $r=2$, it describes an entangling cloner \cite{EntCloner}, i.e., a
beam-splitter of transmission $\tau $\ mixing the signal with one mode of
the TMSV state $\left\vert w\right\rangle $.

Thus, every one-mode Gaussian channel $\mathcal{G}(\mathbf{T},\mathbf{N,d})$
can be uniquely represented by the Stinespring dilation $\{\mathbf{L}(\tau
,r),\left\vert w\right\rangle \}$, up to Gaussian unitaries $\{\mathcal{U}%
_{A},\mathcal{U}_{B}\}$ on the channel and isometries on the environment $%
\mathbf{\tilde{E}}$. By assuming an environment which is bounded in
Euclidean space (i.e., a finite box), the total set of environmental modes
is countable. In such a case, the action of an isometry on $\mathbf{\tilde{E}%
}$ is equivalent to a unitary $\hat{U}_{E}$ involving the two output
ancillas $\mathbf{\tilde{E}}$ and all the remaining ancillas $\mathbf{e}%
=\{e_{i}\}_{i=1}^{\infty }$ of the environment (prepared in the vacuum
state). In other words, $\mathcal{G}(\mathbf{T},\mathbf{N,d})$ can be
represented by the maximal Stinespring dilation $\{\mathbf{L}(\tau ,r)\oplus
\mathbf{I}_{\mathbf{e}},\left\vert w\right\rangle \otimes \left\vert
0\right\rangle _{\mathbf{e}}\}$, up to Gaussian unitaries $\{\mathcal{U}_{A},%
\mathcal{U}_{B}\}$ on the channel and unitaries $\hat{U}_{E}$ on the
environment $\{\mathbf{\tilde{E}},\mathbf{e}\}$ (see Fig.~\ref{PRLpic1},
step~3) \cite{Box}.

Let us now consider the standard cryptographic scenario, where the whole
environment is under control of a malicious eavesdropper (Eve). For each
signal state, Eve can store the corresponding output ancillas $\{\mathbf{%
\tilde{E}},\mathbf{e}\}$ in a quantum memory, detectable by a coherent
measurement $\mathcal{M}_{E}$ at any time of the quantum communication. For
infinite uses of the channel, the output ancillas $\{\mathbf{\tilde{E}},%
\mathbf{e}\}$ will provide an output ensemble of states $\mathcal{E}$. Such
an ensemble can be expressed in terms of Alice's variable $X$ or Bob's
variable $Y$. In other words, there always exist two coherent measurements, $%
\mathcal{M}_{E}(X)$ and $\mathcal{M}_{E}(Y)$, which are optimal in the
estimation of $X$ and $Y$, respectively. This scenario represents the most
general description for a collective Gaussian attack. Luckily, this
description can be greatly simplified if we adopt a suitable
\textquotedblleft entropic bound\textquotedblright\ to restrict Eve's
accessible information on her output ensemble $\mathcal{E}$. This bound can
be provided by the Holevo information, but also by the quantum mutual
information or, more generally, by the von Neumann entropy. On the one hand,
this bound enables us to ignore the details of the quantum measurement $%
\mathcal{M}_{E}$. On the other, since the bound is unitarily invariant, the
environmental unitary $\hat{U}_{E}$ and the extra ancillas \textquotedblleft
$\mathbf{e}$\textquotedblright\ can be also neglected. As a consequence, the
attack's description can be reduced to the set $G:=\{\mathbf{L}(\tau
,r),\left\vert w\right\rangle ,\mathcal{U}_{A},\mathcal{U}_{B}\}$, where $%
\{\tau ,r,w\}$ are the channel symplectic invariants and $\{\mathcal{U}_{A},%
\mathcal{U}_{B}\}$ the input-output Gaussian unitaries (see Fig.~\ref%
{PRLpic1}, step~4). In particular, the Gaussian unitaries $\{\mathcal{U}_{A},%
\mathcal{U}_{B}\}$ are equivalent to a pair of displacements $\{\mathbf{d}%
_{A},\mathbf{d}_{B}\}$ and a pair of symplectic matrices $\{\mathbf{M}_{A},%
\mathbf{M}_{B}\}$. These matrices may be written as $\mathbf{M}_{A}=(\mathbf{%
a}_{1},\mathbf{a}_{2})^{T}$ and $\mathbf{M}_{B}=(\mathbf{b}_{1},\mathbf{b}%
_{2})$, where $\{\mathbf{a}_{1},\mathbf{a}_{2},\mathbf{b}_{1},\mathbf{b}%
_{2}\}$ are $\mathbb{R}^{2}$ column-vectors. The scalar products of these
vectors define three important parameters $\{\theta ,\theta _{A},\theta
_{B}\}$, which contain the basic information about the non-invariant action
of the attack. Explicitly, these parameters are $\theta :=\left\vert \mathbf{%
a}_{1}\right\vert ^{2}\left\vert \mathbf{b}_{1}\right\vert ^{2}+2(\mathbf{a}%
_{1}\cdot \mathbf{a}_{2})(\mathbf{b}_{1}\cdot \mathbf{b}_{2})+\left\vert
\mathbf{a}_{2}\right\vert ^{2}\left\vert \mathbf{b}_{2}\right\vert ^{2}$, $%
\theta _{A}:=\left\vert \mathbf{a}_{1}\right\vert ^{2}+\left\vert \mathbf{a}%
_{2}\right\vert ^{2}$ and $\theta _{B}:=\left\vert \mathbf{b}_{1}\right\vert
^{2}+\left\vert \mathbf{b}_{2}\right\vert ^{2}$. Using the Euler
decomposition \cite{EisertGauss} of the symplectic matrices, we can prove
the lower bounds \cite{Details}%
\begin{equation}
\theta \geq 2~,~\theta _{A}\geq 2~,~\theta _{B}\geq 2~.
\label{MinimaForThetas}
\end{equation}%
Notice that we may call \textquotedblleft canonical\textquotedblright\ the
attacks of the form $C:=\{\mathbf{L}(\tau ,r),\left\vert w\right\rangle ,%
\mathcal{I}_{A},\mathcal{I}_{B}\}$, where $\mathcal{I}$ is the ideal channel
(i.e., the identity map). For this kind of attack it is easy to prove the
minimal condition $\theta =\theta _{A}=\theta _{B}=2$.

Let us now analyze the security of a cvQKD protocol, which is a direct
generalization of the non-switching protocol of Ref.~\cite{Hetero}. In this
protocol, Alice prepares a coherent state $\left\vert \alpha \right\rangle $
whose complex amplitude $\alpha $\ is randomly modulated by a Gaussian
distribution with zero mean and variance $\mu $. Then, Alice sends $%
\left\vert \alpha \right\rangle $ to Bob, who decodes a conditional
amplitude $\beta |\alpha $ by heterodyne detection. Such a process is
repeated many times, with Bob getting an output random amplitude $\beta $
(see Fig.~\ref{PRLpic1}, step~5). At the end of the quantum communication,
part of the data $\{\alpha ,\beta \}$ is publicly disclosed by Alice and
Bob. This step allows them to realize quantum tomography of the Gaussian
channel $\mathcal{G}(\mathbf{T},\mathbf{N},\mathbf{d})$, which completely
discloses $\mathbf{T},\mathbf{N}$ and $\mathbf{d}$. In fact, from the
analysis of the first and second statistical moments, they can fully
retrieve the two transformations of Eq.~(\ref{Displ_Transf}). Thanks to this
information, Bob is able to process his classical data $\beta $ in order to
make an optimal estimation of Alice's signal $\alpha $. Such a classical
post-processing is equivalent to inverting the displacement transformation
in Eq.~(\ref{Displ_Transf}), which generally involves squeezing and rotation
of the two quadratures. Alternatively, Alice can exploit Eq.~(\ref%
{Displ_Transf}) to process her data $\alpha $ and estimate Bob's variable $%
\beta $. The first situation corresponds to \emph{direct reconciliation},
where $\alpha $ is the reference variable, decoded by Bob with the help of
one-way classical communication (CC) from Alice. By contrast, the second
situation corresponds to\emph{\ reverse reconciliation} \cite{EntCloner},
where $\beta $ is the reference variable, decoded by Alice with the help of
one-way CC from Bob. In both cases, the classical mutual information of
Alice and Bob is given by $I(\alpha :\beta )=H(\beta )-H(\beta |\alpha )$,
where $H(\cdots )$ is the Shannon entropy for bivariate Gaussian variables
\cite{CoverThomas}.

The Gaussian channel $\mathcal{G}(\mathbf{T},\mathbf{N},\mathbf{d})$ between
the users is the effect of a collective Gaussian attack. Bounding Eve with
the Holevo information, this attack can be fully characterized by the set $%
G:=\{\mathbf{L}(\tau ,r),\left\vert w\right\rangle ,\mathcal{U}_{A},\mathcal{%
U}_{B}\}$. In this description, the Holevo information $I(\gamma :E)$ of Eve
on the reference variable $\gamma =\alpha ,\beta $ can be computed from the
restricted set of ancillas $\mathbf{\tilde{E}}$ (see Fig.~\ref{PRLpic1}).
The secret-key rate $R$ of the protocol is then equal to $R=\max
\{0,R(\alpha ),R(\beta )\}$, where $R(\gamma ):=I(\alpha :\beta )-I(\gamma
:E)$ is the rate with respect to Alice's variable $\gamma =\alpha $ (direct
reconciliation) or Bob's variable $\gamma =\beta $ (reverse reconciliation).
Let us consider the asymptotic secret-key rate $R_{\infty }:=\lim_{\mu }R$
that can be reached in the limit of high modulation ($\mu \rightarrow
+\infty $). Here, we consider all the values of the transmission $\tau $
with the exception of $\tau =1$. The asymptotic rate $R_{\infty }$ can be
easily proven to be zero for every $\tau \leq 0$ \cite{Details}. By
contrast, in the positive region $0<\tau \neq 1$, the explicit formula of $%
R_{\infty }$ is extremely hard to compute. For this reason, we provide a
lower bound $B_{\infty }\leq R_{\infty }$ which has the non-trivial
advantage of further simplifying the description of the attack. Therefore,
we only consider the positive range $0<\tau \neq 1$ in the remainder of the
paper. It is easy to prove that the mutual information of Alice and Bob has
the asymptotic expression $\lim_{\mu }I(\alpha :\beta )=\log (\mu /\eta )$,
where
\begin{equation*}
\eta :=\frac{1}{\tau }[1+\tau ^{2}+(1-\tau )^{2}w^{2}+\tau \theta
+\left\vert 1-\tau \right\vert w(\tau \theta _{A}+\theta _{B})]^{1/2}.
\end{equation*}%
The latter quantity $\eta $ represents the \emph{total noise} affecting the
quantum communication. It depends on the two invariants $\{\tau ,w\}$ plus
the three non-invariant parameters $\{\theta ,\theta _{A},\theta _{B}\}$
coming from $\{\mathcal{U}_{A},\mathcal{U}_{B}\}$. Let us now bound the
Holevo information $I(\gamma :E)$ of Eve. In direct reconciliation, $%
I(\alpha :E)$ can be bounded using the condition $\theta _{A}\geq 2$ \cite%
{Details}, while, in reverse reconciliation, $I(\beta :E)$ can be bounded by
the quantum mutual information. As a consequence, we get the following bound
on the secret-key rate $R_{\infty }\geq B_{\infty }:=\max \{0,B_{\infty
}(\alpha ),B_{\infty }(\beta )\}$, where%
\begin{equation}
B_{\infty }(\alpha )=\log \left( \frac{2}{e\left\vert 1-\tau \right\vert
\eta }\right) -g(w)+g\left( \tau +\left\vert 1-\tau \right\vert w\right) ~,
\label{R_DR}
\end{equation}%
and%
\begin{equation}
B_{\infty }(\beta )=\log \left( \frac{2}{e\left\vert 1-\tau \right\vert \tau
\eta }\right) -g(w)~,  \label{R_RR}
\end{equation}%
with $g(x):=[(x+1)/2]\log [(x+1)/2]-[(x-1)/2]\log [(x-1)/2]$. Notice that
these asymptotic rates depend only on the three parameters $\{\tau ,w,\eta
\} $. In other words, the significant information about the Gaussian attack $%
G$ is fully contained in the triplet $\{\tau ,w,\eta \}$, where $\tau $ and $%
w$ are symplectic invariants of the channel, while $\eta $ includes the
non-invariant effect of the input-ouput unitaries $\{\mathcal{U}_{A},%
\mathcal{U}_{B}\}$. Such a triplet is completely known to the honest users
thanks to the tomography of the channel and, therefore, the corresponding
value of $B_{\infty }$ can be easily derived.

It is now interesting to analyze the performances of the canonical attacks
in terms of the asymptotic rate $B_{\infty }$. It is easy to show that, for
fixed invariants $\tau $ and $w$, canonical attacks are the less
perturbative and less powerful attacks. In fact, for a canonical attack, we
have $\theta =\theta _{A}=\theta _{B}=2$, so that the total noise $\eta $
takes the minimum value%
\begin{equation}
\eta =1+\frac{1}{\tau }+\frac{\left\vert 1-\tau \right\vert }{\tau }w:=\eta
_{c}(\tau ,w)~.  \label{Eta_minimum}
\end{equation}%
Then, since $B_{\infty }$ is monotonic in $\eta $ [according to Eqs.~(\ref%
{R_DR}) and~(\ref{R_RR})], the minimization of $\eta $ is equivalent to the
maximization of $B_{\infty }$ (for fixed $\tau $ and $w$). By contrast, we
can easily prove that the canonical attacks are the most powerful Gaussian
attacks for fixed transmission $\tau $ and total noise $\eta $. In other
words, for every Gaussian attack, with triplet $\{\tau ,w,\eta \}$, there
always exists a canonical attack, with triplet $\{\tau ,w^{\prime }\geq
w,\eta \}$, such that $B_{\infty }(\tau ,w^{\prime },\eta )\leq B_{\infty
}(\tau ,w,\eta )$. The proof is very easy. The noise $\eta $ of an arbitrary
Gaussian attack $G$ with $\{\tau ,w,\eta \}$ is minimized by the noise $\eta
_{c}(\tau ,w)$ of a canonical attack $C$ with $\{\tau ,w,\eta _{c}(\tau
,w)\} $. Now, let us increase $w$ while keeping $\tau $ fixed in $\{\tau
,w,\eta _{c}(\tau ,w)\}$. From Eq.~(\ref{Eta_minimum}), we see that $\eta
_{c}(\tau ,w)$ increases in $w$ and, therefore, we can choose a value $%
w^{\prime }\geq w$ such that $\eta _{c}(\tau ,w^{\prime })=\eta $. Then, we
get a new canonical attack $C^{\prime }$ with triplet $\{\tau ,w^{\prime
},\eta \}$. But now, also the two quantities $g(w)$ and $g(w)-g(\tau
+|1-\tau |w)$ are increasing in $w$. Therefore, for fixed $\tau $ and $\eta $%
, the condition $w^{\prime }\geq w$ minimizes the rates of Eqs.~(\ref{R_DR})
and~(\ref{R_RR}), which concludes the proof. By combining the previous
results on the asymptotic rate $B_{\infty }(\tau ,w,\eta )$, we deduce that
canonical attacks can be seen as extremal Gaussian attacks, since they
provide upper bounds for fixed $\{\tau ,w\}$ and lower bounds for fixed $%
\{\tau ,\eta \}$.

In conclusion, we have given a simple and compact description of a
completely general collective Gaussian attack. Using such a
characterization, we have derived the asymptotic secret-key rates that are
reachable by a protocol using coherent states, joint measurements of the
quadratures, and one-way classical communications. In particular, the
secret-key rates can be bounded by relatively simple quantities depending on
three channel parameters only. In terms of these bounds, a particular class
of attacks (canonical attacks) can be considered as extremal. Finally, this
work paves the way for completely general security analyses of cvQKD
protocols, where explicit derivations of secret-key rates can be made
without any assumptions on the eavesdropper's interaction.

We thank F. Caruso, V. Giovannetti and A. Holevo for helpful comments. S.P.
was supported by a Marie Curie Fellowship within the 6th European Community
Framework Programme. S.L. was supported by the W.M. Keck center for extreme
quantum information theory (xQIT).


\begin{thebibliography}{99}
\bibitem{Conditions} In particular, $\mathbf{N}$ and $\mathbf{T}$ must
satisfy the properties $\mathbf{N}^{T}=\mathbf{N}>0$ and $\det \mathbf{N}%
\geq \left( \det \mathbf{T}-1\right) ^{2}$.
\bibitem{Decomposition} For $\tau \neq 1$, one always has $\mathbf{L}(\tau
,r)=\mathbf{M}(\tau ,r)\oplus \mathbf{I}_{E_{2}}$, where the expressions of $%
\mathbf{M}(\tau ,r)$ can be found in Ref.~\cite{HolevoVittorio} but in
different shot-noise units. For $\tau =1$ (additive classical-noise
channel), the decomposition of $\mathbf{L}(1,r)$ heavily depends on the rank
$r$. In fact, one has $\mathbf{L}(1,0)=\mathbf{I}$, $\mathbf{L}(1,1)=\mathbf{%
M}(1,1)\oplus \mathbf{I}_{E_{2}}$ and an irreducible $\mathbf{L}(1,2)$.
\bibitem{Box} Results can be then extended to an infinite box.
\bibitem{Details} Detailed derivations will be presented elsewhere.
\end{thebibliography}
\end{document}